\documentclass[journal]{IEEEtran}

\usepackage{cite}

\usepackage{graphicx}

\usepackage[cmex10]{amsmath}
\usepackage{amssymb}

\usepackage{algorithmic}

\usepackage{color}

\begin{document}

% \fi
%\IEEEpeerreviewmaketitle

\title{Energy-Efficient Resource Allocation for Device-to-Device Underlay Communication}

\author{Feiran~Wang,
		Chen Xu,~\IEEEmembership{Student~Member,~IEEE,}
        Lingyang~Song,~\IEEEmembership{Senior~Member,~IEEE,}
        and~Zhu~Han,~\IEEEmembership{Fellow,~IEEE}
\thanks{Manuscript received December 21, 2013; revised May 23, 2014 and September 22, 2014; accepted December 2, 2014.
This work was supported in part by the National 863 Project
No.2014AA01A701, by the National Natural Science Foundation of China
under Grants 61222104 and U1301255, and by National Science and Technology
Major Project under Grant 2013ZX03003003.}
\thanks{F. Wang is with the Department of Electrical Engineering, Stanford University, Stanford, CA 94305, USA (e-mail: feiran@stanford.edu).}
\thanks{C. Xu and L. Song are with the State Key
Laboratory of Advanced Optical Communication Systems and Networks,
School of Electronics Engineering and Computer Science, Peking University,
Beijing, China (e-mail: chen.xu@pku.edu.cn; lingyang.song@pku.edu.cn).}
\thanks{Z. Han is with the Department of Electrical and Computer Engineering,
University of Houston, Houston, TX 77004, USA (e-mail: zhan2@uh.edu).}}

\maketitle

\begin{abstract}
%\boldmath
Device-to-device (D2D) communication underlaying cellular networks is expected to bring significant benefits for utilizing resources, improving user throughput and extending battery life of user equipments. However, the allocation of radio and power resources to D2D communication needs elaborate coordination, as D2D communication can cause interference to cellular communication. In this paper, we study joint channel and power allocation to improve the energy efficiency of user equipments. To solve the problem efficiently, we introduce an iterative combinatorial auction algorithm, where the D2D users are considered as bidders that compete for channel resources, and the cellular network is treated as the auctioneer. We also analyze important properties of D2D underlay communication, and present numerical simulations to verify the proposed algorithm.
\end{abstract}


\section{Introduction}

\IEEEPARstart{T}{he} increasing demand for higher data rates within the available radio spectrum has stimulated various researches on improving spectrum efficiency.
Local area communications have received much attention, as they can further utilize the radio resources and increase cellular capacity.
Device-to-device (D2D) communication as an underlay to cellular network allows user equipments (UEs) in close proximity to communicate directly without data routing through the evolved NodeB (eNB) \cite{Doppler2009}. These UEs share the same resources with cellular users under the careful control of the cellular network. D2D communication is promising for several benefits. The high link quality enables high data rates and low power consumption \cite{Doppler2009}. Besides, D2D communication can improve the spectral efficiency of the system \cite{Fodor2011}, and achieve efficient load balancing \cite{Liu2014}. Due to these potentials, D2D communication has recently gained extensive attention. D2D underlay communication is being studied in Third Generation Partnership Project (3GPP) for Long Term Evolution Advanced (LTE-A) system \cite{3GPP2012}. Industries have continued to push this study item, and have been working on solutions for D2D communication \cite{Wu2010}. Moreover, D2D communication can contribute to new types of wireless services, such as the concert network \cite{Doppler2009}, relay by smartphone \cite{Nishiyama2014}, and proximity-aware internetworking \cite{Corson2010}.

To incorporate D2D underlay communication into cellular networks, some modification is needed, e.g., D2D session setup management, peer discovery, and physical layer procedures need to be designed \cite{Doppler2009a,Fodor2012a,Hakola2010}.
Another challenge is that D2D communication can cause interference due to sharing cellular resources. Consequently, the system needs efficient coordination of cellular and D2D communication to manage the communication quality. Various allocation techniques have been proposed to cope with these problems \cite{Fodor2011, Yu2009a, Yu2011, Yu2009, Min2011a, Zulhasnine2010, Belleschi2011, Xu2012, Janis2009, Xing2010}.
In \cite{Yu2009a,Yu2011}, the optimal resource sharing mode and power control were discussed under the spectral efficiency restrictions.
In \cite{Yu2009}, the authors applied a power control method which constrains the the signal to interference plus noise ratio (SINR) degradation of the cellular link to a certain level.
In \cite{Min2011a}, the closed form expressions of outage probability for three different receive modes were derived.
These works mainly focused on only one D2D link.
The problem of radio resource allocation to D2D communication in a network was formulated as a mixed integer non-linear programming in \cite{Zulhasnine2010}, and a greedy heuristic was proposed to solve the problem.
In \cite{Belleschi2011}, the authors proposed a distributed suboptimal joint mode selection and resource allocation scheme.
In \cite{Xu2012}, the authors proposed a sequential second price auction mechanism to allocate the spectrum resources for D2D communications.
As is proved in these work, by proper coordination, interference can be limited and D2D communication can greatly improve the performance of the network.

In contrast to existing work, in this paper we consider joint power and radio resource allocation for D2D communication, and energy efficiency is considered as our optimization objective since the devices are handheld equipments with limited battery capacity. The energy consumption of UEs includes transmission energy and circuit energy.
The circuit energy consumption is the energy consumed by the circuit blocks along the signal path. To characterize the non-linear effects in battery, we employ 
the Peukert's law \cite{Rao2003} to model the battery lifetime.
Note that this paper allows multiple D2D pairs sharing the same resource. This can further utilize the spectrum, while prior work typically assumes that one channel resource can only be reused by
one D2D pair. We introduce a combinatorial auction (CA) game to solve the allocation problem.
Combinatorial auctions are multi-item or multi-bidder auctions in which bidders can form combinations called packages, rather than just bid individually~\cite{Cramton2006}. The concept of package in CA corresponds to multiple D2D sharing the same channel. Furthermore, the valuation of resource allocation depends on the combinatorial performance of the UEs rather than individual UEs, which corresponds to the valuations in CA. Hence, the resource allocation problem can be formulated as a CA problem correspondingly.
In an iterative combinatorial auction, the steps of bid evaluation are executed multiple times, allowing bidders to better express their valuations \cite{Pikovsky2008}.
CAs have already found applications in allocating radio spectrum for wireless communications \cite{Pikovsky2008,Xu2012a,Pal2007}.
The main problem and challenge concerning CAs is the winner determination problem (WDP), which needs careful consideration in designing the auction mechanism \cite{Vries2003,Andersson2000, Sandholm2002,Wang2012}.

In our CA game, the D2D pairs are viewed as bidders, while the cellular networks are sellers. The game has two nested levels, the channel allocation level and the power control level. In the outer level, each bidder has a utility function for the channels, and multiple bidders can form a package that share the same channel. The inner power control level is modeled as a non-cooperative game. Some game-theoretic power control schemes were studied in \cite{Saraydar2002, Meshkati2006, Wang2013a}, and were shown to have good performances in terms of energy efficiency. The co-channel UEs are viewed as non-cooperative players. They compete with each other and adjust their power to maximize their individual payoffs. During the auction, the bidders submit bids and the seller decides the allocation of the channels.
The auction runs iteratively until reaching an equilibrium state. We design the mechanism of the CA game to solve the resource allocation problem, and investigate some important properties of the game, e.g., existence and uniqueness of the equilibrium, and convergence. We also investigate how the performance of UEs changes
with the number of D2D pairs and maximum D2D communication distance.

The contributions of this paper are summarized as follows.
\begin{itemize}
\item We propose to optimize the expected data during battery lifetime, instead of traditional energy efficiency. We characterize the non-linear effects in battery by using an empirical model (Peukert's Law) for battery lifetime. The proposed metric is a more practical indicator of energy efficiency. To the best of our knowledge, previous works have not considered such an objective.
\item We study joint channel and power allocation for D2D communication and we allow multiple D2D users sharing a channel. We formulate this problem as a combinatorial auction, where the two problems have close correspondence. 
\item We propose a novel algorithm for solving the resource allocation problem. We design a nested two-level game. The outer level deals with channel allocation, and the inner level deals with power control. The two levels are jointly performed, rather than in a sequential order. We also discussed the properties of our algorithm, and the influence of system parameters rigorously.
\end{itemize}

The rest of the paper is organized as follows. In Section \ref{sec_system}, we briefly introduce the system model of the D2D communication underlaying cellular network. Next, in Section \ref{sec_prob}, we formulate the resource allocation problem as a CA game. We develop the two-level resource allocation, and analyze important properties of the proposed game. In Section \ref{sec_sim}, we present the simulation results. Finally, we conclude this paper in Section \ref{sec_conclusion}.

\section{System Model\label{sec_system}}

We consider a single cell environment with multiple UEs and one eNB located at the center of the cell. The UEs and the eNB are equipped with a single omni-directional antenna. The system contains two types of UEs, cellular UEs and D2D UEs, where we assume that their communication modes have been selected. The cellular UEs communicate through the eNB, while the D2D UEs communicate directly to each other in pairs, and the two UEs in a pair are close to each other.
The number of cellular UEs and D2D pairs in the system are $K$ and $D$ ($D<K$), respectively. There are $K$ orthogonal channels, and each cellular UE occupies an orthogonal channel. The D2D UEs can reuse the channel resources occupied by the cellular UEs, and multiple D2D pairs can share the same channel simultaneously.

\begin{figure}[!t]
\centering
\includegraphics[width=3in]{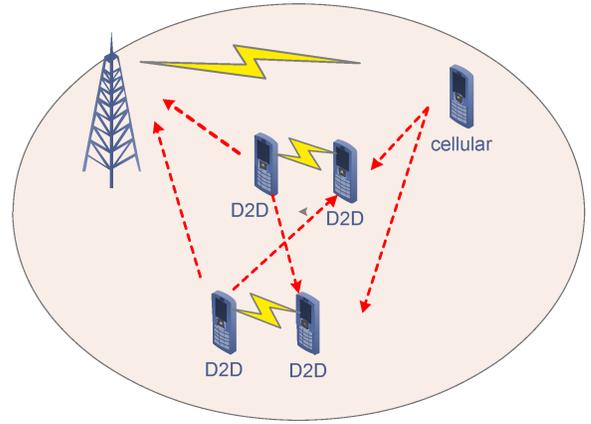}
\caption{System model of D2D underlay communication with uplink resource sharing. UE$_1$ is a cellular UE whereas UE$_2$ and UE$_3$, UE$_4$ and UE$_5$ are in D2D communication. Directed lines indicate interference.}
\label{fig:sys}
\end{figure}

D2D session setup procedures are described in \cite{Doppler2009}. We assume the eNB has perfect channel state information (CSI) of all channels. The eNB allocates radio resources to D2D UEs to make the system work efficiently. Moreover, the cellular and D2D UEs adjust transmit power to coordinate the interference between them. We also assume that the inter-cell interference is managed efficiently with the inter-cell interference control mechanism \cite{Min2011a}. Hence, we only consider the intra-cell interference.

A scenario of uplink resource sharing is illustrated in Fig.~\ref{fig:sys} where one cellular UE (UE$_1$) and two D2D pairs (UE$_2$ and UE$_3$, UE$_4$ and UE$_5$) are sharing the same channel. UE$_2$ and UE$_4$ are transmitters while UE$_3$ and UE$_5$ are receivers. The two D2D UEs in a D2D pair are close enough to satisfy the maximum distance constraint of D2D communication, in order to guarantee the quality of D2D service. During the uplink period of the cellular network, the cellular UE (UE$_1$) transmits data to the eNB, while the eNB suffers interference from D2D transmitters (UE$_2$ and UE$_4$). The D2D pairs are in communication while the receivers (UE$_3$, UE$_5$) are exposed to interference from the cellular UE (UE$_1$) and the other D2D transmitters (UE$_4$, UE$_2$, respectively).

During the downlink period, the cellular UEs receive data from the eNB, and interference from D2D UEs sharing the same channel. D2D UEs suffer serious interference from the eNB, due to the large transmit power of eNB compared to UEs, making it hard to guarantee the quality of D2D services. Therefore, we focus on uplink frame of the network.

As illustrated above, during the uplink phase of the cellular network, the eNB suffers interference from D2D pairs sharing the resources. If D2D pair $d$ reuses the channel of the $k$-th cellular UE, it receives interference from cellular UE $k$ and other transmitters of D2D pair $d'(d'\neq d)$ sharing the same channel.
We define $\mathcal{D}_k$ as a set of variables representing the indexes of D2D pairs that share the channel with the $k$-th cellular UE. Each set is called a package. The channel rate of $k$-th cellular UE is calculated as
\begin{equation}
  r_k = \log_2\left(1 + \dfrac{p_k g_{ke}}{\sum\limits_{d\in\mathcal{D}_k} p_d g_{de} + \sigma^2}\right).
  \label{eq: rate_k}
\end{equation}
The channel rate of D2D pair $d$ ($d\in\mathcal{D}_k$) is
\begin{equation}
  r_d = \log_2\left(1 + \dfrac{p_d g_{dd}}{p_k g_{kd}+\sum\limits_{d'\in\mathcal{D}_k\backslash{d}} p_{d'} g_{d'd} + \sigma^2}\right),
  \label{eq: rate_d}
\end{equation}
where $p_c$, $p_d$ and $p_{d'}$ are the transmit power of cellular UE $k$ and D2D pair $d$, $d'$, respectively. $g_{ke}$ denotes the channel gain between the $k$-th cellular UE and the eNB. Similarly, $g_{de}$ represents the channel gain between the $d$-th D2D pair and the eNB. $g_{ij}$ denotes the channel gain between the $i$-th cellular UE or D2D pair and the $j$-th cellular UE or D2D pair. $\sigma^2$ accounts for the thermal noise power at the receiver. $\sum_{d\in\mathcal{D}_k} p_d g_{de}$ in (\ref{eq: rate_k}) is the received interference power at the eNB from D2D pairs in $\mathcal{D}_k$. $p_k g_{kd} + \sum_{d'\in\mathcal{D}_k\backslash{d}} p_{d'} g_{d'd}$ in (\ref{eq: rate_d}) is the interference power from cellular UE $k$ and D2D pairs $d'\in\mathcal{D}_k, d' \neq d$.
The system sum rate during the uplink period can be expressed as
\begin{equation}
  r_t=\sum_{k=1}^{K} r_k+ \sum_{d=1}^D r_d.
\end{equation}

In this paper, the expected battery lifetime is considered as an important optimization parameter. The energy consumption of D2D UEs includes two parts, the transmission energy and the circuit energy. The circuit energy is the energy consumed by all the circuit blocks along the signal path \cite{Cui2004}. It cannot be ignored since it has an important effect on the battery lifetime. Without loss of generality, we assume all UEs have the same constant circuit power consumption $p_0$.
Note that, the energy drawn from a battery is not always equivalent to the energy consumed in device circuits \cite{Rao2003}. To capture the non-linear effect, we model the battery lifetime $l$ using Peukert's law\begin{equation}
  l=\frac{C}{I^a},
\end{equation}
where $C$ is the battery capacity. $I$ is the discharge current. $a$ is a constant around 1.3.
For UE$_i$ with transmit power $p_i$ and operating voltage $V_0$, the expected battery lifetime $l_i$ is
\begin{equation}
  l_i=\frac{CV_0^a}{(p_i+p_0)^a}.
\end{equation}

Here is a fundamental tradeoff for each UE between battery lifetime and transmission rate. The transmit power needs to be properly controlled to balance between the two aspects. Therefore, we propose to maximize the expected quantity of data transmission during the battery lifetime, i.e.,
\begin{equation}
  u_i = r_i l_i,
  \label{eq: utility}
\end{equation}
which is also defined as the utility function for each UE. This is a metric for energy efficiency, since energy efficiency is the ratio of the quantity of data to the energy consumption. 
The motivation for using expected data over traditional definition of energy efficiency (transmission rate over power consumption) is that this metric takes into account the non-linear effects of batteries. Rather than the traditional instantaneous energy efficiency, this metric characterizes the expected data during battery lifetime, which is more practical, and better in capturing energy efficiency.
Note that this is a measure of expected total data from one time instance, not the calculation of actual data transmitted. 

\section{Combinatorial Resource Auction\label{sec_prob}}

In this paper, energy efficiency is explicitly considered, which is critical since the UEs are handheld equipments with limited battery lifetime. We assume the channels of the cellular UEs have been already allocated, while we study the allocation radio resource to D2D UEs. In addition, the transmit power of the cellular and the D2D UEs are jointly adjusted to mitigate the interference in the network.
We develop a two-level combinatorial auction game, corresponding to channel allocation and power control. The two levels are jointly played. In the channel allocation level, D2D UEs are viewed as the bidders while the eNB is the seller. Since multiple D2D pairs can share one channel, the packages of D2D pairs can form combinatorial bidders. The seller sells channel resources to maximize their combinatorial utilities. Here, the combinatorial utility for a channel refers to the sum of the utilities of UEs on that channel. In the power control level, the co-channel UEs are players, and they compete through power control to maximize their own utilities. Through the interaction of players, an equilibrium state can be reached.

\subsection{Channel Allocation Level\label{sec: channelAllocation}}
In the system, there are $D$ bidders (D2D pairs) and $K$ items (cellular UEs and corresponding channels). The D2D pairs are combined to form D2D packages $\mathcal{D}_k, k=1,2,...,K$ in correspondence with the $k$-th cellular UE. The possible packages are subsets of D2D pairs. The combinatorial utility for the package is defined as the payoffs of all the UEs on the channel, i.e.,
\begin{equation}
  U_k = u_k + \sum_{d\in\mathcal{D}_k} u_d.
\end{equation}
The utilities of the bidders can also be viewed as the virtual money the seller receives. For each bidder package, the goal is to maximize its combinatorial utility. The combinatorial resource auction can be formulated as
\begin{equation}
\begin{split}
  \max \quad & U_k, \forall k\\
  \mathrm{s.t.} \quad & \mathcal{D}_i \cap \mathcal{D}_j = \emptyset, i,j = 1,2,...,K, i \neq j\\
  & \bigcup_{k=1}^{K} \mathcal{D}_k = \{1,2,...,D\}.
\end{split}
\end{equation}

Here, the first constraint ensures that a D2D pair can only be in one package. The second constraint guarantees that each D2D pair can be allocated one channel. This problem is called the combinatorial allocation problem in CA games, also referred to as the winner determination problem \cite{Pikovsky2008}. For the combinatorial auction, it has been proved that no polynomial-time algorithm can be constructed for achieving the reasonable worst case guarantee \cite{Sandholm2002}, i.e., the problem is NP hard. We therefore propose an approximate solution to solve the problem in a tractable manner.

We use a multi-round iterative combinatorial auction. In the first round, the bidders calculate utilities and submit bids for each channel. The bids are equal to the gains of the combinatorial utilities compared to the initial utilities. The seller finds the highest bid pair $(k^*, d^*)$, and sells the channel occupied by the ${k^*}$-th cellular UE to bidder $d^*$. $d^*$ will be added to $\mathcal{D}_{k^*}$. Then, every bidder recalculates utilities for the allocated channel. We allow the channel to be sold to another bidder, but the bidder cannot bid for more channels. The auction process moves on until all the bidders obtain an item. Then, the auction enters the second round. In the second round, the seller adjusts the auction results to improve the outcome. The seller tries to remove the bidder $d$ ($d = 1,2,...,D$) from package $\mathcal{D}_k$ ($k = 1,2,...,K$), and finds the bidder $d^\dag \in \mathcal{D}_{k^\dag}$ that the combinatorial utility increase the most, and kicks $d^\dag$ from $\mathcal{D}_{k^\dag}$. The kicked bidder bids again for other channels, and $d^\dag$ is put into $\mathcal{D}_{k^\dag}$ that has the largest increase in utility. If the total increase from the adjustment is positive, the seller is willing to sell the channel. Otherwise, $d^\dag$ is put back into $\mathcal{D}_{k^\dag}$. The above process repeats until the packages cannot be adjusted. We also set that each D2D pair can only be adjusted once.

\subsection{Power Control Level}
\newtheorem{definition}{Definition}
\newtheorem{proposition}{Proposition}

When calculating utilities for bidder $d$ and package $\mathcal{D}_k$, cellular UE $k$ and D2D pairs in $\mathcal{D}_k$ adjust power to maximize their individual utilities. All the co-channel UEs are viewed as players in a non-cooperative power control game. The number of players is $N$. For simplicity, we denote the players using index $i=1,2,...,N$. Each player selects a transmit power $p_i \in [0,\bar{p}]$, where $\bar{p}$ is the maximum transmit power for each UE. The power vector $\textbf{p} = (p_1, p_2, ..., p_N)^T$ denotes the outcome of the game. The utility function of player $i$ is defined as the expected data in (\ref{eq: utility}). We denote the utility function alternatively as $u_i(\mathbf{p})$ or $u_i(p_i, \mathbf{p}_{-i})$, where $\mathbf{p}_{-i}$ represents the power of players excluding $i$. The game can be expressed as
\begin{equation}
  \max u_i(p_i,\mathbf{p}_{-i}), \forall i.
  \label{eq:power_obj}
\end{equation}

The utility of each user depends on its transmit power and also on other players' strategies. Given other players' strategies, the player can choose a transmit power that maximizes its own utility. The strategy constitutes a best response function.
%\begin{definition}
%The best response $b_i(\mathbf{p}_{-i})$ for player $i$ is
%\begin{equation}
%  b_i(\mathbf{p}_{-i}) = \{ \left. p_i\in[0,\bar{p}] \right| u_i(p_i,\mathbf{p}_{-i})\geq u_i(p'_i,\mathbf{p}_{-i}),\forall p_i' \in[0,\bar{p}] \}.
%\end{equation}
%\end{definition}
We also refer to the best response as the optimal power $p_i^*$.
\begin{proposition}
  The best response for player $i$ is
  \begin{equation}
    b_i(\mathbf{p}_{-i}) = \min(\tilde{p}_i,\bar{p}),
  \end{equation}
  where $\tilde{p}_i$ is the maximum point in $(0, \infty)$, i.e., $\tilde{p}_i= \arg\max_{p_i\in\mathbb{R}^+} u_i(p_i,\mathbf{p}_{-i}).$
\end{proposition}

The proof can be found in the appendix.
In the game-theoretic scenario, the players make decisions and interact with each other. It is necessary to solve for an equilibrium state for the game. A widely used solution is Nash equilibrium \cite{Osborne1994}.
%\begin{definition}
%A set of strategies $\mathbf{p}$ is a Nash equilibrium if no unilateral deviation %in strategy by any single player is profitable for that player, i.e.,
%\begin{equation}
%  u_i(\mathbf{p}_i,\mathbf{p}_{-i})\geq u_i(\mathbf{p}_i',\mathbf{p}_{-i}),\forall i.
%\end{equation}
%\end{definition}
A Nash equilibrium offers a stable outcome of a game in which multiple selfish players compete through self-optimization and reach a point where no player wishes to deviate. From another perspective, a Nash equilibrium is the strategies that are the best responses for all players, i.e. $p_i \in b_i(\mathbf{p}_{-i}),\forall i$. We have the following conclusion on the existence and uniqueness of Nash equilibrium in the power control game.

\begin{proposition}
A Nash equilibrium exists in the power control game.
\end{proposition}

\begin{proposition}
  The power control game has a unique equilibrium if
  \begin{equation}
    p_0\tilde{p}_i + \frac{I_i - \sigma^2}{g_{ii}} > 0, \forall i.
  \end{equation}
\end{proposition}

The proofs can be found in the appendix. In a Nash equilibrium, every player achieves optimal power. Player $i$'s utility achieved at the equilibrium state is
%\begin{equation}
%  u_i^*(\mathbf{p}^*) =
%  \begin{cases}
%    \dfrac{CV_0^a}{a\ln 2} \dfrac{\alpha_i}{(1+\tilde{p}_i\alpha_i)(\tilde{p}_i+p_0)^{a-1}} \  & \text{if} \  {p}^*_{i} = \tilde{p}_i;\\[1em]
%    \dfrac{CV_0^a}{\ln 2} \dfrac{\ln(1+\bar{p}\alpha_i)}{(\bar{p}+p_0)^a} \  & \text{if} \  {p}^*_{i} = \bar{p}.
%  \end{cases}
%  \label{eq: optimalUtility}
%\end{equation}
Since Nash equilibrium exists and is unique, the equilibrium can be found using an iterative algorithm according to the fixed point theory. We set a sequence of power $p_i^{n},n=0,1,2,...$, and update as follows
\begin{enumerate}
  \item initialize $n=0$, $p_i^0=0, \forall i$, and $\epsilon>0$;
  \item update the power sequence using the best response function $p_i^{n+1} = b_i(\mathbf{p}_{-i}^n), \forall i$;
  \item if $|p_i^{n+1}-p_i^n|<\epsilon, \forall i$, terminate; else, goto 2).
\end{enumerate}

\begin{figure}[!t]
\centering\small
\algsetup{indent=2em}
\hrule
\vskip 0.3em
Algorithm 1. Joint Channel and Power Allocation Algorithm\\
\vskip 0.3em
\hrule
\vskip 0.4em

\renewcommand\baselinestretch{1}\selectfont
\begin{algorithmic}[1]
\STATE Initialize $\mathcal{D}_1,\mathcal{D}_2,...,\mathcal{D}_{K}$ to be empty set;
\STATE Setup $\epsilon>0$, $\mathbf{U}^0_{K\times D}$ the initial utility matrix, $\mathbf{U}_{K\times D}$ the utility calculation matrix, $\mathbf{P}_{K\times D}$ the power control matrix;

\FOR{$d=1$ to $D$}
  \FOR{$k=1$ to $K$}
    \STATE Calculate optimal power if the $d$-th D2D pair shares the $k$-th cellular UE's channel: $\mathbf{P}_{kd}=b_d(p_k)$, using the iterative method in Section III-B;
    \STATE Calculate utilities $u_k$, $u_d$, $\mathbf{U}_{kd}=u_k+u_d - \mathbf{U}^0_{kd}$;
  \ENDFOR
\ENDFOR

\FOR{$d=1$ to $D$}
  \STATE Find the maximum utility element $(k^*,d^*)$;
  \STATE Put $d^*$ into $\mathcal{D}_{k^*}$: $\mathcal{D}_{k^*} = \mathcal{D}_{k^*}\cup\{d^*\}$;
  \STATE Set $\mathbf{U}_{\cdot d^*}=\mathbf{0}$;
  \STATE Update other players’ optimal power and utilities for the $k^*$-th UE's channel;
\ENDFOR

\WHILE{\TRUE}
  \STATE Initialize $\delta_0=0$;
  \FOR{$d=1$ to $D$}
    \STATE If $d$ has been adjusted, skip;
    \STATE Find the package $k$ that $d$ is in;
    \STATE Try to remove $d$ from package $k$, and calculate utility $U_k'$;
    \IF{$U_k'-U_k>\delta_0$}
      \STATE $d^*=d$, $\delta_0=U_k'-U_k$;
    \ENDIF
  \ENDFOR

  \STATE Mark $d^*$;
  \STATE Set $\delta_1=0$;
  \FOR{$k=1$ to $K$}
    \STATE Try to add $d^*$ to package $k$, and calculate utility $U_k'$;
    \IF{$U_k'-U_k>\delta_1$}
      \STATE $k^*=k$, $\delta_1=U_k'-U_k$;
    \ENDIF
  \ENDFOR

  \IF{$\delta_0 + \delta_1 > 0$}
    \STATE Put $d^*$ into $\mathcal{D}_{k^*}$;
  \ELSE
    \STATE \textbf{break};
  \ENDIF

\ENDWHILE
\end{algorithmic}
%\caption{Resource and power allocation algorithm}
\label{fig:algorithm}
\vskip 0.3em
\hrule
\end{figure}

\subsection{Overall mechanism}

The overall mechanism includes the above two nested levels to jointly allocate channel and power. The outer level is described in Section~\ref{sec: channelAllocation}, and for every step in the channel allocation level, the power control game is played to determine the optimal power. Specifically, when adjusting packages, we use the iterative method in Section III-B to and calculate optimal power and corresponding utilities. The overall algorithm is summarized in Algorithm~1.

\section{Analysis of the CA algorithm}
In this section, we analyze computational complexity of our algorithm, and the signaling overhead to incorporate our algorithm into practical networks. We also analyze the effect of number of UEs and maximum D2D distance on system performances.

\subsection{Convergence and Computational Complexity}
In this subsection, we investigate the convergence and computational complexity of the proposed algorithm. We first study the convergence of the iterative power control algorithm. The following definition provides us with a description of the speed of convergence \cite{Burden2011}.
\begin{definition}
  Suppose $\{\mathbf{p}^{n}\}_{n=1}^\infty$ is a sequence that converges to $\mathbf{p}^*$, if positive constants $q$ and $\lambda$ exist with
  \begin{equation}
    \lim_{n\rightarrow\infty} \frac{\|\mathbf{p}^{n+1} - \mathbf{p}^*\|}{\|\mathbf{p}^{n} - \mathbf{p}^*\|^q} = \lambda,
  \end{equation}
  then $\{\mathbf{p}^{n}\}_{n=1}^\infty$ converges to $\lambda$ of order $q$. Moreover, if $q = 1$ and $\lambda > 0$, the sequence converges linearly.
\end{definition}

\begin{proposition}
  The power control game converges linearly to the unique Nash equilibrium.
\end{proposition}
\begin{IEEEproof}
We first prove that the power iteration sequence is increasing by induction. The initial point is $\mathbf{p}^0=0$, and $\mathbf{p}^1=\mathbf{b}(\mathbf{p}^0)>0=\mathbf{p}^0$. Suppose $\mathbf{p}^{n'}>\mathbf{p}^{n'-1}$ holds for $n'=n,n-1,...,2$. For $n'=n+1$, we have $\mathbf{p}^{n+1}=\mathbf{b}(\mathbf{p}^n)\geq \mathbf{b}(\mathbf{p}^{n-1})=\mathbf{p}^n$, where the inequality holds from the fact that $\mathbf{b}(\mathbf{p})$ is monotonically increasing in $\mathbf{p}$. Thus, the power iteration sequence is an increasing sequence. Furthermore, the sequence has an upper bound $\bar{p}$. The sequence converges since an increasing sequence with an upper bound has a limit \cite{Burden2011}.

Let $\lim_{n\rightarrow\infty}p_i^n=p_i^*$. At any iteration $n$, we can find a sequence $p_i'^n\in[0,\bar{p}]$ such that $\lim_{n\rightarrow\infty}p_i'^n=p_i'^*$. Since $p_i^n$ is the best response of player $i$, we have
\begin{equation}
  u_i(p_i^n,\mathbf{p}_{-i}^n) \geq u_i(p_i'^n,\mathbf{p}_{-i}'^n),
\end{equation}
and let $n$ approach infinity,
\begin{equation}
  u_i(p_i^*,\mathbf{p}_{-i}^*) \geq u_i(p_i'^*,\mathbf{p}_{-i}'^*).
\end{equation}
The inequity holds for all $i$. Hence, $p^*$ is an equilibrium.

If $\frac{\partial\mathbf{b}}{\partial\mathbf{p}}\neq 0$ near $\tilde{\mathbf{p}}$, the iteration converges linearly \cite{Burden2011}.
We proved $\frac{\partial\mathbf{b}}{\partial\mathbf{p}}\neq 0$ in (\ref{eq: deravativeToPower}). Therefore, the iteration linearly converges to the unique equilibrium.
\end{IEEEproof}

\begin{proposition}
The channel and power allocation algorithm based on CA concludes in finite time.
\end{proposition}
\begin{IEEEproof}
  In the first round, $D$ D2D pairs are allocated sequentially. In the second round, when a D2D pair is adjusted from one package to another package, the combinatorial utilities are nondecreasing. Since there are a finite number of combinations, this round takes finite time.
\end{IEEEproof}

In the first round of the proposed algorithm, every channel is evaluated for all D2D pairs, resulting in a computation of $O(KD)$. In the second round, since each D2D pair can only be adjusted for no more than once, the complexity is $\sum_{j=1}^D (D-j)K = O(KD^2)$. Thus, the complexity of the proposed algorithm is $O(KD^2)$. Here, for a given function $h(n)$, we denote by $O(h(n))$ the set of functions $O(h(n))=\{f(n):\exists\  c,n_0 \ \text{such that} \ 0\leq f(n) \leq ch(n), \ \text{for all}\  n\geq n_0\}$.
We compare our algorithm to the greedy heuristic in \cite{Zulhasnine2010}. The procedures of the greedy heuristic is briefly described as follows:
\begin{enumerate}
  \item Sort queue of signal-to-noise ratio (SNR) for all uplink UEs in decreasing order;
  \item While not all D2D UEs allocated;
  \item Dequeue one channel, which has the largest SNR in the queue;
  \item Allocate the channel to the D2D pair for which channel gain is minimum.
\end{enumerate}
The complexity for the greedy heuristic is $O(KD)$. Our algorithm has a higher complexity, but has a much better performance, which is shown in Section \ref{sec_sim}.

\subsection{Signaling Overhead}
Obtaining CSI between UEs and the eNB requires no additional signaling overhead compared to existing resource allocation schemes, such as proportional fairness and maximum carrier to interference. The additional information needed in our scheme is CSI between D2D and cellular UEs, and CSI between D2D pair UEs. At the beginning of the allocation, the D2D transmitters send detection signals to estimate CSI. The estimated CSI is then reported to the eNB. The following iteration process is all conducted at the eNB, and no signal overhead is needed in the network until the control signal forwarding. Also, channel adaption methods such as \cite{Conti2007,Li2010} can be incorporated into our resource allocation scheme to adapt to large-scale channel variations, further reducing signal overhead. But this paper focused on near-optimal solutions, and exploring various channel adaption techniques is part of our future work.

\subsection{Number of UEs\label{sec: number}}
When the number of UEs in the network changes, the performance of each UE changes accordingly. When the number of D2D pairs in the network grows while the number of cellular UEs remains unchanged, the utilities of the cellular UEs will decrease. We prove this result as follows. Suppose there are $D$ D2D pairs in the system, and a new D2D pair $d'$ comes in, channel allocation for $d'$ has two cases:
\begin{enumerate}
  \item $d'$ reuses the channel occupied by the $k$-th cellular UE, and the channel is not shared by other D2D pairs;
  \item $d'$ shares a channel with other D2D pairs.
\end{enumerate}
For the first case, the $k$-th cellular UE receives the newly imposed interference by D2D pair $d'$, and thus the channel quality $\alpha_k$ degrades. The equilibrium utility can be viewed as a function of the channel quality for each UE. We take the total derivative of the utility function at $p_k^*$ with respect to $\alpha_k$, and get
\begin{equation}
  \left. \frac{d u_k}{d \alpha_k} \right|_{p_k=p_k^*}
  = \left.\frac{\partial u_k}{\partial \alpha_k}\right|_{p_k=p_k^*} + \left.\frac{\partial u_k}{\partial p_k}\right|_{p_k=p_k^*}  \frac{\partial p_k^*}{\partial \alpha_k}.
\end{equation}
We can easily verify that $\frac{\partial u_k}{\partial \alpha_k} = CV_0^a p_k/[\ln2 (1+p_k\alpha_k)(p_k+p_0)^a]>0, \forall p_k\in[0,\bar{p}]$. If $p_k^* =\tilde{p}_k$, from the first order optimality condition we have $\left.\frac{\partial u_k}{\partial p_k}\right|_{p_k=p_k^*} = 0$. Thus, $\left. \frac{d u_k}{d \alpha_k} \right|_{p_k=p_k^*}>0$. If $p_k^* =\bar{p}$, we have $\frac{\partial p_k^*}{\partial \alpha_k} = 0$. We also derive $\left. \frac{d u_k}{d \alpha_k} \right|_{p_k=p_k^*}>0$. This indicates that the equilibrium utility for the cellular UE is monotonically decreasing with $\alpha_k$. Consequently, the entering of D2D pair $d'$ will make the cellular UE worse off.

For the second case, suppose package $\mathcal{D}_k$ has players $i=1,2,...,N$, and the equilibrium utilities are $u_i^*, i=1,2,...,N$. After D2D pair $d'$ joins package $\mathcal{D}_k$, the equilibrium utilities are $u_i'^*, i=1,2,...,N$ and $u_{d'}'^*$. We are ready to prove that $u_i'^*<u_i^*, \forall i$, which means that if a new D2D pair enters a non-empty package, the utility of every player in that package will decrease.
In the new equilibrium $p_i'^*, i=1,2,...,N, p_{d'}^*$, we can view the interference from D2D pair $d'$ as a part of noise, i.e., $\sigma_i'^2 = \sigma^2 + p_{d'}^*g_{d'i}, i=1,2,...,N$, and have $\sigma_i'^2 > \sigma^2$. Moreover, we can prove that the equilibrium utility is monotonically decreasing with $\sigma^2$, since
\begin{equation}
  \left. \frac{\partial u_k}{\partial \sigma^2} \right|_{p_k=p_k^*}
  = \left. \frac{\partial u_k}{\partial \alpha_k} \right|_{p_k=p_k^*} \frac{\partial \alpha_k}{\partial \sigma^2}
  < 0.
\end{equation}
Therefore, we have $u_i'^*<u_i^*, \forall i$, which finishes the proof. This indicates that the entering of a new D2D pair makes every UE on the channel worse off. However, the new D2D pair will gain from getting access to the channel. The converse also holds that the leaving of a D2D pair from a package will make the left players better off. We can also infer that with more D2D pairs, the average performance for each D2D pair will remain nearly the same, since the performance of D2D communication depends more on the proximity of UEs rather than the channel resources. On the other hand, if the number of channels increase, the D2D pairs have more channels to choose from, resulting in an improvement of D2D pairs' average performances.

\subsection{Maximum D2D Communication Distance\label{sec: distance}}
In D2D underlaying cellular network, the maximum communication distance between two D2D UEs is a crucial parameter. In network design, it can be used as a criterion for the eNB to decide whether to set up a direct link between the two UEs. If the distance between two UEs are within the threshold, they can communicate directly. The main consideration here is to utilize the advantage of proximity between UEs, which enables higher data rates and lower power consumption. On the contrary, D2D needs more transmit power for a large distance, causing more interference to the cellular network. To show this, we take the derivative of optimal power with respect to distance $d_{ii}$ between the D2D UEs in pair $i$, and derive
\begin{equation}
  \frac{\partial \tilde{p}_i}{\partial d_{ii}}
  = -\frac{\partial f_i}{\partial d_{ii}}/{\frac{\partial f_i}{\partial \tilde{p}_i}}
  = -\frac{\partial f_i}{\partial \alpha_i} \frac{\partial \alpha_i}{\partial g_{ii}} \frac{\partial g_{ii}}{\partial d_{ii}} / {\frac{\partial f_i}{\partial \tilde{p}_i}}
  > 0.
\end{equation}
The inequality holds since we have $\frac{\partial f_i}{\partial \alpha_i}<0$ from (\ref{eq: derivativeToAlpha}), $\frac{\partial g_{ii}}{\partial d_{ii}} < 0$, $\frac{\partial \alpha_i}{\partial g_{ii}} = 1/(I_i + \sigma^2) > 0$, and $\frac{\partial f_i}{\partial \tilde{p}_i} < 0$ from (\ref{eq: deravativeToMyPower}). This shows that with a larger communication distance between D2D UEs, the optimal transmit power increase, in order to overcome the signal attenuation.

\subsection{Fairness}
Historical information can be used to ensure the fairness of D2D users. We proposed in \cite{Wang2013} to add one penalty term to the utility function indicating the cumulative utilities. This can be easily incorporated into our algorithm. In this paper, we focus on maximizing energy efficiency.

\section{Simulation Results\label{sec_sim}}
In this section, we provide several simulation results to evaluate the performance of D2D communication and the proposed algorithm.
A single circular cell environment is considered. The cellular UEs and D2D pairs are uniformly distributed in the cell. The transmitter and receiver in a D2D pair are close enough to satisfy the distance requirement of D2D communication. We focus on the uplink period of the system. The results are averaged over 1000 realizations.
%The channel is modeled as Rayleigh fading channel and the free-space path loss model is used. The channel gain consists of path loss and the Rayleigh fading coefficient.
We assume that all channel coefficients follow the independent complex Gaussian distribution, and free space propagation pathloss model is used. The received signal power is
\begin{equation}
  p_r=p_t d_{tr}^{-2}|h_{tr}|^2,
\end{equation}
where $p_r$ and $p_t$ are the received power and the transmit power, respectively. $d_{tr}$ is the distance between transmitter $t$ and receiver $r$. $h_{tr}$ represents the complex Gaussian channel coefficient that satisfies $h_{tr}\sim\mathcal{CN}(0,1)$.
Main simulation parameters are presented in \tablename~\ref{table:simulation}.

\begin{table}[!t]
%\renewcommand{\arraystretch}{1.3}
\caption{Simulation Parameters and Values}
\label{table:simulation}
\centering
% Some packages, such as MDW tools, offer better commands for making tables
% than the plain  tabular which is used here.
\begin{tabular}{ll}%{p{1.6in}p{1.55in}}
\hline
Parameters & Values\\
\hline
Cellular layout & one isolated circular cell\\
Cell radius & 350m\\
UE distribution & randomly distributed\\
Number of cellular UEs & 30\\
Number of channel resources & 30\\
Number of D2D pairs & 6-30, 4\\
Maximum UE Tx power & 200mW (23dBm)\\
UE antenna gain & 0dBi\\
Channel bandwidth & 180kHz\\
Thermal noise power density & -174dBm/Hz\\
Circuit power consumption & 50mW (17dBm)\\
Battery capacity & 800mA$\cdot$h \\
Operating voltage & 4V \\
$\epsilon$ & 1mW\\
Realizations & 1000\\
\hline
\end{tabular}
\end{table}

\subsection{Convergence}
\begin{figure}[!t]
\centering
\includegraphics[width=3in]{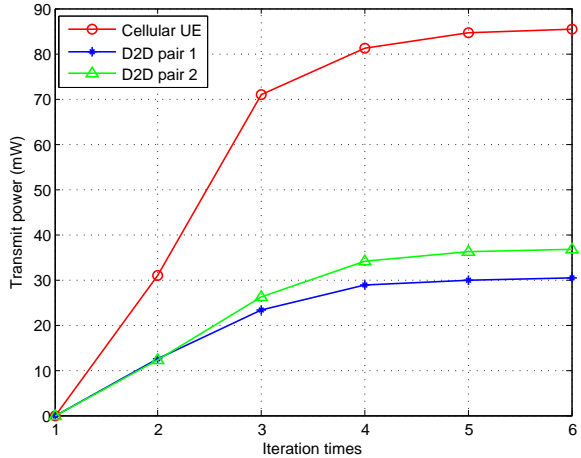}
\caption{Transmit power with iteration times.}
\label{fig: power_iteration}
\end{figure}

We give one realization of the system, and plot the transmit power of UEs on one channel with the iteration times in Fig.~\ref{fig: power_iteration}. There are one cellular UE and two D2D pairs using this channel. Starting from 0, the transmit power is adjusted according to the iterative power control algorithm in Section~\ref{sec_prob}. The stopping parameter is $\epsilon = 1$mW. We can see that the transmit power converges to the equilibrium in 6 iterations. Parameter $\epsilon$ controls the speed of convergence. For a larger $\epsilon$, the power control algorithm converges faster, but with larger error, and vice versa.

\subsection{Performance of Different Algorithms}

\begin{figure}[!t]
\centering
\includegraphics[width=3.5in]{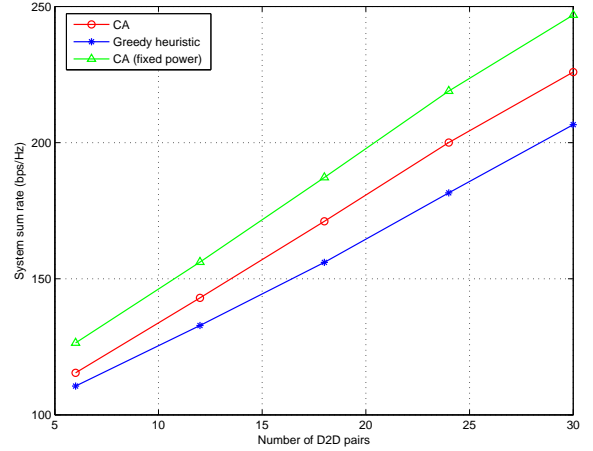}
\caption{System sum rate with number of D2D pairs.}
\label{fig: sumrate_number}
\end{figure}

As a comparison, we compare the performance of our algorithm to the greedy heuristic in \cite{Zulhasnine2010}.
Since the greedy heuristic only deals with channel allocation, we incorporate our power control algorithm into the greedy heuristic. Specifically, when a D2D pair is allocated, the transmit power of UEs on the channel is controlled using our power control algorithm. We also investigate the performance of our algorithm with fixed transmit power at 50mW.

\begin{figure}[!t]
\centering
\includegraphics[width=3.5in]{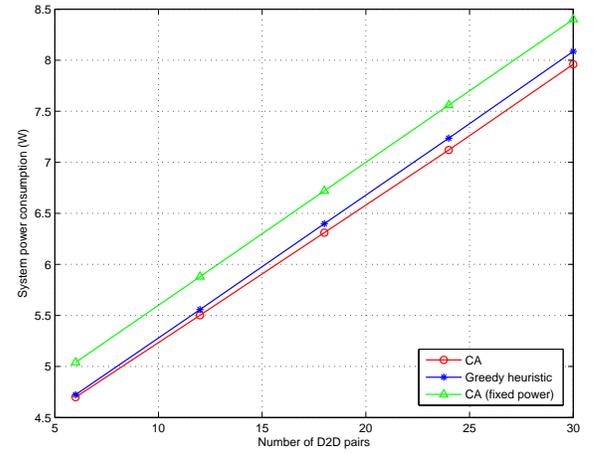}
\caption{System power consumption with number of D2D pairs.}
\label{fig: power_number}
\end{figure}

\begin{figure}[!t]
\centering
\includegraphics[width=3.5in]{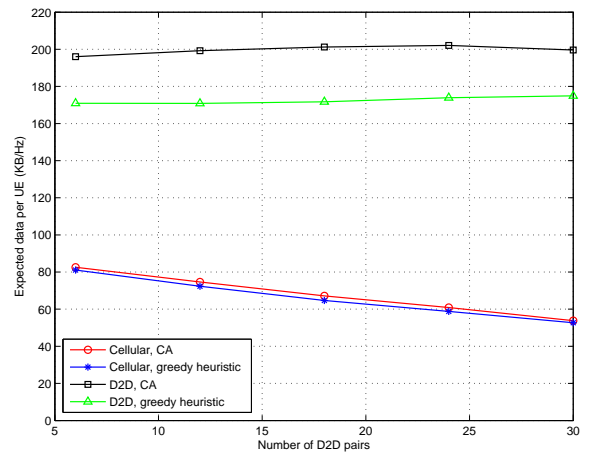}
\caption{Expected data per UE with number of D2D pairs.}
\label{fig: dataPerUE_number}
\end{figure}

We first study the system-level performances. We plot the sum rate of the system with different numbers of D2D pairs in Fig.~\ref{fig: sumrate_number}.
We can see that all curves go up with more D2D pairs, which indicates that with the joining of D2D communication, the sum rate of the network is increased.
This is due to the proximity of D2D pairs such that UEs can achieve high data rates.
Our algorithm behaves better than the greedy heuristic, and the absolute performance gap expands with more D2D pairs. The relative performance gap when there are 6 D2D pairs are about 10\%.
CA with fixed power achieves a higher sum rate since it consumes more energy, which is illustrated in Fig.~\ref{fig: power_number}.
Fig.~\ref{fig: power_number} shows the sum power consumption of the system with different numbers of D2D pairs. With more D2D pairs entering the network, the power consumption of the system increases dramatically. (Our algorithm consumes the about the same energy as the greedy heuristic since they use the same power control scheme.)

\begin{figure}[!t]
\centering
\includegraphics[width=3.5in]{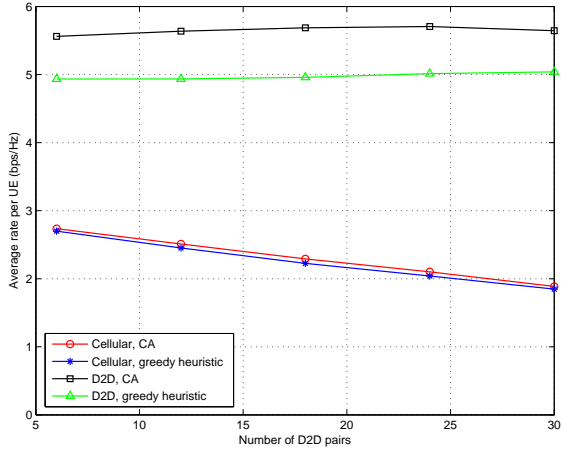}
\caption{Average rate per UE with number of D2D pairs.}
\label{fig: ratePerUE_number}
\end{figure}

\begin{figure}[!t]
\centering
\includegraphics[width=3.5in]{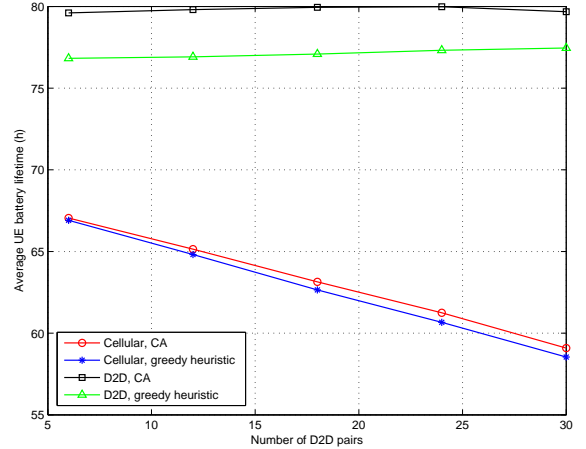}
\caption{Average UE battery lifetime with number of D2D pairs.}
\label{fig: lifetime_number}
\end{figure}

We then study the performance of each UE. In Fig.~\ref{fig: dataPerUE_number} -- Fig.~\ref{fig: lifetime_number}, we plot the expected data during battery lifetime per UE, average rate per UE and average battery lifetime with different numbers of D2D pairs, respectively. It is obvious that D2D communication performs much better than cellular communication, about 150\% larger in expected data transmission, 100\% higher in data rates, and 25\% longer in battery lifetime.
With more D2D pairs entering the network, the performances of each D2D UE nearly remain unchanged, but the performances of each cellular UE degrade rapidly. This is due to the fact that the newly entered D2D pairs bring interference to the cellular UEs, and is consistent with the analysis in Section \ref{sec: number}.

\begin{figure}[!t]
\centering
\includegraphics[width=3.1in]{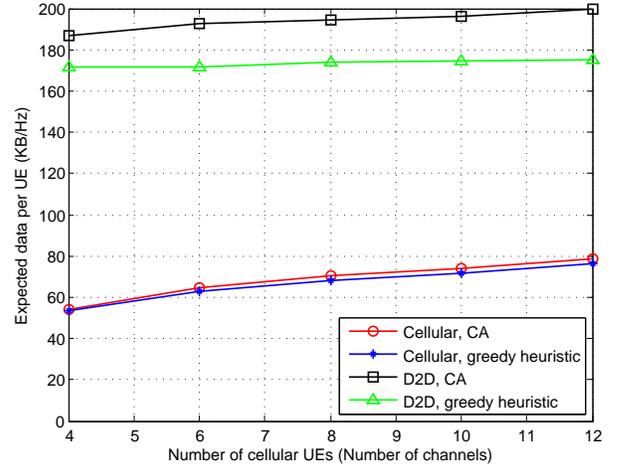}
\caption{Expected data per UE with number of channels (cellular UEs).}
\label{fig: dataPerUE_cell}
\end{figure}

In Fig.~\ref{fig: dataPerUE_cell}, we plot system energy efficiency with the number of channels. The number of D2D pairs is 4. Note that we assume the number of cellular UEs equals the number of channels, since each channel is occupied by a cellular UE. We observe that with the increasing number of channels, the expected data for both the cellular and D2D UEs increases. With more channels, D2D pairs have more resources to choose from, and thus the performance is improved. However, the increase is slow, which implies that the cellular UEs do not have major impact on the performance of D2D UEs. On the contrary, the performance improvement of cellular UEs is more significant. With more cellular UEs and channels, the average interference per cellular UE receives is lower, resulting in a performance gain. We can also infer that when the number of channels continues to increase, the performance of UEs will not change dramatically.

\subsection{Performance with Different D2D Distances}
\begin{figure}[!t]
\centering
\includegraphics[width=3.2in]{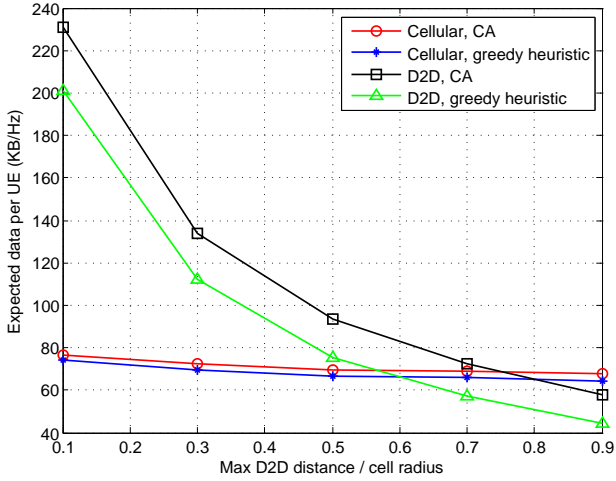}
\caption{Expected data per UE for different maximum D2D communication distances.}
\label{fig: dataPerUE_distance}
\end{figure}

\begin{figure}[!t]
\centering
\includegraphics[width=3.2in]{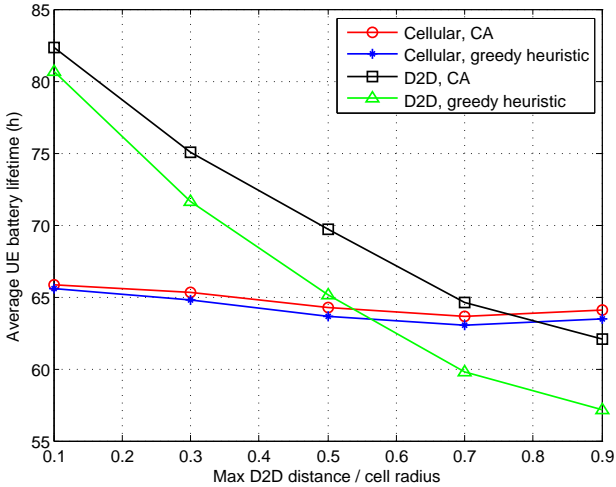}
\caption{Average rate per UE for different maximum D2D communication distances.}
\label{fig: lifetime_distance}
\end{figure}

To study the effects of the maximum D2D communication distance on system performance, we plot the UE performances with different maximum D2D distances in Fig.~\ref{fig: dataPerUE_distance} and Fig.~\ref{fig: lifetime_distance}, where we use the ratio of maximum D2D distance to the cell radius as the metric.

Fig.~\ref{fig: dataPerUE_distance} illustrates the effects of maximum D2D communication distance on the expected data. The expected data per D2D UE decreases dramatically with a larger maximum D2D distance, about 40\% per 0.1 cellular radius. With the increasing distance between two D2D UEs in the pair, D2D communication gradually lose its benefit of proximity. This corresponds to the analytical results in Section \ref{sec: distance}. When the maximum D2D distance equals about 0.8 of the cell radius, D2D has the same expected data as the cellular UEs for our algorithm (about 0.6 for the greedy heuristic).

For a larger D2D communication distance, D2D UEs need more transmit power to guarantee the quality of service, resulting in more energy consumption and shorter battery lifetime. This is shown in Fig.~\ref{fig: lifetime_distance}. The average D2D UE battery lifetime shortens 2.5h with 0.1 cell radius increase in the maximum D2D distance. When the maximum D2D distance exceeds about 0.8 of the cell radius, D2D UEs have lower battery lifetime than the cellular UEs.

The results show that limiting the maximum communication distance of D2D can effectively improve the system performances in terms of battery lifetime and expected data. However, in practical network, restricting the distance means a smaller number of UEs can use D2D mode. Consequently, there is a tradeoff between the performance of D2D communication and the number of UEs using D2D mode. The maximum D2D distance should be properly designed for the practical systems. Generally, it should at least guarantee that D2D communication performs better than the cellular UEs.

\section{Conclusion\label{sec_conclusion}}
In this paper, we investigated joint channel and power allocation for device-to-device communication underlaying cellular network. We formulated the optimization problem as a combinatorial auction, and proposed a joint channel and power allocation algorithm, in order to improve the energy efficiency for each UE. The proposed algorithm can converge in finite rounds. 
Simulation results show that the proposed algorithm achieves better performances than the greedy heuristic and the fixed power algorithm in terms of expected data, lifetime and data rate. In addition, the results show that by using the proposed resource allocation algorithm, D2D communication underlaying cellular networks can increase the system energy efficiency compared to the pure cellular system.  Moreover, Both analytical and simulated indicate that the number of D2D pairs and maximum D2D communication distance have negative impacts on the performance of individual UEs.

% or


\appendices

\section{Proof of Best Response}
\label{appendix:response}
For simplicity, we denote the channel rate of player $i$ as $r_i = \log_2(1 + p_i \alpha_i)$, where $\alpha_i = g_{ii}/(I_i+\sigma^2)$ and can be viewed as the effective channel quality, $g_{ii}$ is the channel gain between player $i$ and the receiver communicating with $i$, and $I_i = \sum_{j\neq i}p_j g_{ji}$ is the interference received at the receiver of the $i$-th player. Note that the channel quality $\alpha_i$ only depends on other players' transmit power $\mathbf{p}_{-i}$. We take the partial derivative of $u_i$ with respect to $p_i$ and derive
\begin{equation}
\begin{split}
  \frac{\partial u_i}{\partial p_i} &= 
   \frac{CV_0^a}{(p_i+p_0)^{a+1}\ln 2} \left[\frac{(p_i+p_0)\alpha_i}{1+p_i\alpha_i} - a\ln(1+p_i\alpha_i)\right] \\&
   \begin{cases}
    = 0 \quad p_i\in(0,\bar{p});\\
    > 0 \quad p_i = \bar{p}.
  \end{cases}.
\end{split}
\end{equation}
Let $f_i = (p_i+p_0)\alpha_i/(1+p_i\alpha_i) - a\ln(1+p_i\alpha_i)$. The above is from evaluating $f_i$ at $p_i=0$ gives $p_0\alpha_i > 0$, and ${CV_0^a}/[(p_i+p_0)^{a+1}\ln 2]>0, \forall p_i\in[0,\bar{p}]$.
%\begin{equation}
%  f_i  \begin{cases}
%    = 0 \quad p_i\in(0,\bar{p});\\
%    > 0 \quad p_i = \bar{p}.
%  \end{cases}
%  \label{eq: condition}
%\end{equation}
Taking the partial derivative of $f_i$ with respect to $p_i$, we have
\begin{equation}
  \frac{\partial f_i}{\partial p_i} = \alpha_i\frac{1-p_0\alpha_i-a(1+p_i\alpha_i)}{(1+p_i\alpha_i)^2} <0, \forall p_i>0,
  \label{eq: deravativeToMyPower}
\end{equation}
since $a>1$. This indicates that $f_i$ is monotonically decreasing with $p_i$. We already observed $f_i(0) = p_0\alpha_i$, and we also notice that $f_i \rightarrow -\infty$ as $p_i\rightarrow\infty$. Thus, according to the mean value theorem, there exists a unique point $\tilde{p}_i>0$ such that $f_i=0$, and the point is the global maximum point for all $p_i>0$. Considering the boundaries of the feasible set, we conclude that the best response is given by $b_i(\mathbf{p}_i) = \min(\tilde{p}_i,\bar{p})$.

\section{Proof of existence of Nash equilibrium}
\label{appendix:existence}
A Nash equilibrium exists \cite{Osborne1994}, if $\forall i$
\begin{enumerate}
  \item the set of strategies is a nonempty compact convex subset of a Euclidean space;
  \item the utility function is continuous and quasi-concave.
\end{enumerate}

Obviously, the set of player $i$'s strategies $[0,\bar{p}]$ is a nonempty compact convex subset of $\mathbb{R}$, and the utility function $u_i(p_i,\mathbf{p}_{-i})$ is continuous in $p_{i}$.

Function $u_i$ is quasi-concave if for all $p_i, p_i'\in[0,\bar{p}]$ and $\lambda\in[0,1]$, we have
\begin{equation}
  u_i(\lambda p_i + (1 - \lambda)p_i',\mathbf{p}_{-i}) \geq \min\left(u_i(p_i,\mathbf{p}_{-i}), u_i(p_i',\mathbf{p}_{-i})\right).
\end{equation}
Without loss of generality, we suppose $p_i<p_i'$. When $\bar{p}\leq\tilde{p}_i$, $f_i>0,\forall p_i\in[0,\bar{p}]$. $u_i$ is monotonically increasing in $[0,\bar{p}]$. Thus, $u_i(\lambda p_i + (1 - \lambda)p_i',\mathbf{p}_{-i})\geq u_i(p_i,\mathbf{p}_{-i})$.  When $\bar{p}>\tilde{p}_i$, we discuss the following cases:
\begin{enumerate}
  \item If $p_i'\leq \tilde{p}_i$, we have $f_i > 0, \forall p_i \in [0, \tilde{p}_i]$, and $u_i$ is monotonically increasing in $[0,\tilde{p}_i]$. Thus, $u_i(\lambda p_i + (1 - \lambda)p_i',\mathbf{p}_{-i})\geq u_i(p_i,\mathbf{p}_{-i})$.
  \item If $p_i\geq \tilde{p}_i$, similarly we can derive $u_i(\lambda p_i + (1 - \lambda)p_i',\mathbf{p}_{-i})\geq u_i(p_i',\mathbf{p}_{-i})$.
  \item If $p_i < \tilde{p}_i < p_i'$, let $\tilde{\lambda}_i = (p_i'-\tilde{p}_i)/(p_i'-p_i)$. We have $u_i(\lambda p_i + (1 - \lambda)p_i',\mathbf{p}_{-i})\geq u_i(p_i,\mathbf{p}_{-i})$ when $\lambda < \tilde{\lambda}_i$; $u_i(\lambda p_i + (1 - \lambda)p_i',\mathbf{p}_{-i})\geq u_i(p_i',\mathbf{p}_{-i})$ when $\lambda \geq \tilde{\lambda}_i$.
\end{enumerate}
Consequently, $u_i(p_i,\mathbf{p}_i)$ is quasi-concave in $p_i$.

\section{Proof of uniqueness of Nash equilibrium}
\label{appendix:uniqueness}
The Nash equilibrium satisfies $\mathbf{p} = \mathbf{b}(\mathbf{p})$, where $\mathbf{b}(\mathbf{p})=(b_1(\mathbf{p}),b_2(\mathbf{p}),...,b_N(\mathbf{p}))$. We know that the fixed point $\mathbf{p} = \mathbf{b}(\mathbf{p})$ is unique for a standard function \cite{Saraydar2002} defined below
\begin{definition}
A function is standard if it satisfies
\begin{itemize}
  \item $\mathbf{b(p)}>0$;
  \item if $\mathbf{p}\geq\mathbf{p'}$, then $\mathbf{b(p)}\geq \mathbf{b(p')}$;
  \item $\forall \mu>1, \mu\mathbf{b(p)}>\mathbf{b}(\mu\mathbf{p})$.
\end{itemize}
Note the inequality of vector means that every component satisfies the inequality.
\end{definition}

The best response function is always positive, i.e., $b_i(\mathbf{p}_{-i})>0, \forall i$. For the second property, by taking the partial derivative of $f_i$ at $\tilde{p}_i$ with respect to $\alpha_i$ and considering $f_i(\tilde{p}_i)=0$, we get
%\begin{equation}
%  \left.\frac{\partial f_i}{\partial\alpha_i} \right|_{p_i = \tilde{p}_i} = \frac{\tilde{p}_i+p_0-a\tilde{p}_i(1+\tilde{p}_i\alpha_i)}{(1 + \tilde{p}_i\alpha_i)^2}.
%  \label{eq: derivativeToAlpha}
%\end{equation}
%From $f_i(\tilde{p}_i)=0$, we have
%\begin{equation}
%  \tilde{p}_i+p_0 = \frac{1 + \tilde{p}_i\alpha_i}{\alpha_i} a \ln(1 + \tilde{p}_i\alpha_i).
%\end{equation}
%Substituting the above equation into (\ref{eq: derivativeToAlpha}), we obtain
\begin{equation}
  \left.\frac{\partial f_i}{\partial\alpha_i} \right|_{p_i = \tilde{p}_i} = \frac{a}{1+\tilde{p}_i\alpha_i} \left[ \frac{1}{\alpha_i}\ln(1+\tilde{p}_i\alpha_i) - \tilde{p}_i \right] < 0.\label{eq: derivativeToAlpha}
\end{equation}
The inequality follows since $\frac{1}{\alpha_i}\ln(1+\tilde{p}_i\alpha_i) - \tilde{p}_i < 0$ for all $\tilde{p}_i>0$. Then, we take the partial derivative of $\tilde{p}_i$ with respect to $p_j,\forall j\neq i$. Note $f_i=0$ constructs an implicit function of $\tilde{p}_i$ and $p_j,\forall j\neq i$, and derive
\begin{equation}
\begin{split}
  \frac{\partial \tilde{p}_i}{\partial p_j}
 = -\frac{\partial f_i}{\partial \alpha_i} \frac{\partial \alpha_i}{\partial p_j} / {\frac{\partial f_i}{\partial \tilde{p}_i}}
  > 0.
\end{split}
\label{eq: deravativeToPower}
\end{equation}
The inequality holds since we have $\frac{\partial f_i}{\partial \tilde{p}_i}<0$ from (\ref{eq: deravativeToMyPower}), and $\frac{\partial \alpha_i}{\partial p_j} = -\alpha_i^2 g_{ji}/g_{ii} < 0$.
We obtain that $\tilde{p}_i$ is monotonically increasing with $p_j,\forall j\neq i$.
Thus, the best response function $b_i(\mathbf{p}_{-i})=\min(\tilde{p}_i,\bar{p})$ is monotonically increasing with $\mathbf{p}_{-i}$, for all $i$. This indicates that when other players use larger transmit power, the optimal power for player $i$ also increases to overcome the inference. We can also derive similarly ${\partial \tilde{p}_i}/{\partial \alpha_i}<0$, which means that with a better channel quality, the optimal power is lower.

For the third property, let $\alpha_i' = {g_{ii}}/({\mu I_i + \sigma^2})$, and $f_i'=f_i|_{\alpha_i=\alpha_i'}$. Since $f_i'$ is monotonically decreasing in $p_j, \forall j\neq i$, we apply $f_i'$ to both sides of $\mu\mathbf{b(p)}>\mathbf{b}(\mu\mathbf{p})$ and get $f_i'(\mu \tilde{p}_i)<0$. To prove $\mu\mathbf{b(p)}>\mathbf{b}(\mu\mathbf{p})$), it is equivalent to prove $f_i'(\mu \tilde{p}_i)<0$. We have
\begin{equation}
  f_i'(\mu \tilde{p}_i) = \frac{(\mu\tilde{p}_i+p_0)\alpha_i'}{1+\mu\tilde{p}_i\alpha_i'} - a\ln(1+\mu\tilde{p}_i\alpha_i').
\end{equation}
Since $\alpha'>\alpha/\mu$, $a\ln(1+\mu\tilde{p}_i\alpha_i') > a\ln(1+\tilde{p}_i\alpha_i)$. Considering $f_i(\tilde{p}_i)=0$, to prove $f_i'(\mu \tilde{p}_i)<0$, we only need to prove ${(\mu\tilde{p}_i+p_0)\alpha_i'}/{(1+\mu\tilde{p}_i\alpha_i')} < {(\tilde{p}_i+p_0)\alpha_i}/{(1+\tilde{p}_i\alpha_i)}$. We have
\begin{equation}
  \frac{(\mu\tilde{p}_i+p_0)\alpha_i'}{1+\mu\tilde{p}_i\alpha_i'} - \frac{(\tilde{p}_i+p_0)\alpha_i}{1+\tilde{p}_i\alpha_i} = \frac{(1-\mu)[p_0\tilde{p}_i+(I_i-\sigma^2)/g_{ii}]}{ (\mu\tilde{p}_i+1/{\alpha_i'})(\tilde{p}_i+1/\alpha_i)}.
\end{equation}
Since $\mu>1$, we
derive $f_i'(\mu \tilde{p}_i)<0$ if $p_0\tilde{p}_i+(I_i-\sigma^2)/g_{ii}>0$.

%\section{}

%\section*{Acknowledgment}


\ifCLASSOPTIONcaptionsoff
  \newpage
\fi



\bibliographystyle{IEEEtran}
\bibliography{IEEEabrv,d2d}

\begin{thebibliography}{10}
\providecommand{\url}[1]{#1}
\csname url@samestyle\endcsname
\providecommand{\newblock}{\relax}
\providecommand{\bibinfo}[2]{#2}
\providecommand{\BIBentrySTDinterwordspacing}{\spaceskip=0pt\relax}
\providecommand{\BIBentryALTinterwordstretchfactor}{4}
\providecommand{\BIBentryALTinterwordspacing}{\spaceskip=\fontdimen2\font plus
\BIBentryALTinterwordstretchfactor\fontdimen3\font minus
  \fontdimen4\font\relax}
\providecommand{\BIBforeignlanguage}[2]{{%
\expandafter\ifx\csname l@#1\endcsname\relax
\typeout{** WARNING: IEEEtran.bst: No hyphenation pattern has been}%
\typeout{** loaded for the language `#1'. Using the pattern for}%
\typeout{** the default language instead.}%
\else
\language=\csname l@#1\endcsname
\fi
#2}}
\providecommand{\BIBdecl}{\relax}
\BIBdecl

\bibitem{Doppler2009}
K.~Doppler, M.~Rinne, C.~Wijting, C.~Ribeiro, and K.~Hugl, ``Device-to-device
  communication as an underlay to {LTE}-advanced networks,'' \emph{{IEEE}
  Commun. Mag.}, vol.~47, no.~12, pp. 42--49, Dec. 2009.

\bibitem{Fodor2011}
G.~Fodor and N.~Reider, ``A distributed power control scheme for cellular
  network assisted {D2D} communications,'' in \emph{Proc. IEEE Global
  Telecommunications Conference}, Houston, TX, Dec. 2011.

\bibitem{Liu2014}
J.~Liu, Y.~Kawamoto, H.~Nishiyama, N.~Kato, and N.~Kadowaki, ``Device-to-device
  communications achieve efficient load balancing in lte-advanced networks,''
  \emph{{IEEE} Wireless Commun. Mag.}, vol.~21, no.~2, pp. 57--65, Apr. 2014.

\bibitem{3GPP2012}
{3GPP, RP-122009}, ``Study on {LTE} device to device proximity services - radio
  aspects,'' 2012.

\bibitem{Wu2010}
X.~Wu, S.~Tavildar, S.~Shakkottai, T.~Richardson, J.~Li, R.~Laroia, and
  A.~Jovicic, ``{FlashLinQ}: a synchronous distributed scheduler for
  peer-to-peer ad hoc networks,'' in \emph{Proc. 48th Annual Allerton
  Conference on Communication, Control, and Computing}, Allerton, IL, Sep.
  2010, pp. 514--521.

\bibitem{Nishiyama2014}
H.~Nishiyama, M.~Ito, and N.~Kato, ``Relay-by-smartphone: realizing multihop
  device-to-device communications,'' \emph{IEEE Commun. Mag.}, vol.~52, no.~4,
  pp. 56--65, Apr. 2014.

\bibitem{Corson2010}
M.~S. Corson, R.~Laroia, J.~Li, V.~Park, T.~Richardson, and G.~Tsirtsis,
  ``Toward proximity-aware internetworking,'' \emph{{IEEE} Wireless Commun.
  Mag.}, vol.~17, no.~6, pp. 26--33, Dec. 2010.

\bibitem{Doppler2009a}
K.~Doppler, M.~P. Rinne, P.~Janis, C.~Ribeiro, and K.~Hugl, ``Device-to-device
  communications; functional prospects for lte-advanced networks,'' in
  \emph{Proc. IEEE International Conference on Communications Workshops},
  Dresden, Germany, Jun. 2009.

\bibitem{Fodor2012a}
G.~Fodor, E.~Dahlman, G.~Mildh, S.~Parkvall, N.~Reider, G.~Miklos, and
  Z.~Turanyi, ``Design aspects of network assisted device-to-device
  communications,'' \emph{{IEEE} Commun. Mag.}, vol.~50, no.~3, pp. 170--177,
  Mar. 2012.

\bibitem{Hakola2010}
S.~Hakola, T.~Chen, J.~Lehtoma, and T.~Koskela, ``Device-to-device
  communication in cellular network - performance analysis of optimum and
  practical communication mode selection,'' in \emph{Proc. IEEE Wireless
  Communications and Networking Conference}, Sydney, Australia, Apr. 2010.

\bibitem{Yu2009a}
C.-H. Yu, O.~Tirkkonen, K.~Doppler, and C.~Ribeiro, ``Power optimization of
  device-to-device communication underlaying cellular communication,'' in
  \emph{Proc. IEEE International Conference on Communications}, Dresden,
  Germany, Jun. 2009.

\bibitem{Yu2011}
C.-H. Yu, K.~Doppler, C.~B. Ribeiro, and O.~Tirkkonen, ``Resource sharing
  optimization for device-to-device communication underlaying cellular
  networks,'' \emph{{IEEE} Trans. Wireless Commun.}, vol.~10, no.~8, pp.
  2752--2763, Aug. 2011.

\bibitem{Yu2009}
C.-H. Yu, O.~Tirkkonen, K.~Doppler, and C.~Ribeiro, ``On the performance of
  device-to-device underlay communication with simple power control,'' in
  \emph{Proc. IEEE 69th Vehicular Technology Conference Spring}, Barcelona,
  Spain, Apr. 2009.

\bibitem{Min2011a}
H.~Min, W.~Seo, J.~Lee, S.~Park, and D.~Hong, ``Reliability improvement using
  receive mode selection in the device-to-device uplink period underlaying
  cellular networks,'' \emph{{IEEE} Trans. Wireless Commun.}, vol.~10, no.~2,
  pp. 413--418, Feb. 2011.

\bibitem{Zulhasnine2010}
M.~Zulhasnine, C.~Huang, and A.~Srinivasan, ``Efficient resource allocation for
  device-to-device communication underlaying {LTE} network,'' in \emph{Proc.
  IEEE International Wireless and Mobile Computing, Networking and
  Communications Conference}, Niagara Falls, NJ, Oct. 2010, pp. 368--375.

\bibitem{Belleschi2011}
M.~Belleschi, G.~Fodor, and A.~Abrardo, ``Performance analysis of a distributed
  resource allocation scheme for {D2D} communications,'' in \emph{Proc. IEEE
  Global Communications Conference Workshops}, Houston, TX, Dec. 2011, pp.
  358--362.

\bibitem{Xu2012}
C.~Xu, L.~Song, Z.~Han, Q.~Zhao, X.~Wang, and B.~Jiao, ``Interference-aware
  resource allocation for device-to-device communications as an underlay using
  sequential second price auction,'' in \emph{Proc. IEEE International
  Conference on Communications}, Ottawa, Canada, Jun. 2012.

\bibitem{Janis2009}
P.~Janis, V.~Koivunen, C.~Ribeiro, J.~Korhonen, K.~Doppler, and K.~Hugl,
  ``Interference-aware resource allocation for device-to-device radio
  underlaying cellular networks,'' in \emph{Proc. IEEE 69th Vehicular
  Technology Conference Spring}, Barcelona, Spain, Sep. 2009.

\bibitem{Xing2010}
H.~Xing and S.~Hakola, ``The investigation of power control schemes for a
  device-to-device communication integrated into {OFDMA} cellular system,'' in
  \emph{Proc. IEEE 21st International Personal Indoor and Mobile Radio
  Communications Symposium}, Instanbul, Turkey, Sep. 2010, pp. 1775--1780.

\bibitem{Rao2003}
R.~Rao, S.~Vrudhula, and D.~N. Rakhmatov, ``Battery modeling for energy aware
  system design,'' \emph{Computer}, vol.~36, no.~12, pp. 77--87, Dec. 2003.

\bibitem{Cramton2006}
P.~Cramton, Y.~Shoham, and R.~Steinberg, \emph{Combinatorial Auctions}.\hskip
  1em plus 0.5em minus 0.4em\relax MIT Press, Cambridge, MA, 2006.

\bibitem{Pikovsky2008}
A.~Pikovsky, ``Pricing and bidding strategies in iterative combinatorial
  auctions,'' Ph.D. dissertation, Technischen Universitat Munchen, 2008.

\bibitem{Xu2012a}
C.~Xu, L.~Song, Z.~Han, D.~Li, and B.~Jiao, ``Resource allocation using a
  reverse iterative combinatorial auction for device-to-device underlay
  cellular networks,'' in \emph{Proc. IEEE Global Communications Conference},
  Anaheim, CA, Dec. 2012, pp. 4542--4547.

\bibitem{Pal2007}
S.~Pal, S.~Kundu, M.~Chatterjee, and S.~Das, ``Combinatorial reverse auction
  based scheduling in multi-rate wireless systemshao,'' \emph{{IEEE} Trans.
  Comput.}, vol.~56, no.~10, pp. 1329 --1341, Oct. 2007.

\bibitem{Vries2003}
S.~de~Vries and R.~Vohra, ``Combinatorial auctions: a survey,'' \emph{Informs
  Journal On Computing}, vol.~15, no.~3, pp. 284--309, Summer 2003.

\bibitem{Andersson2000}
A.~Andersson, M.~Tenhunen, and F.~Ygge, ``Integer programming for combinatorial
  auction winner determination,'' in \emph{Proc. Fourth International
  Conference on MultiAgent Systems}, Boston, MA, Jul. 2000, pp. 39--46.

\bibitem{Sandholm2002}
T.~Sandholm, ``Algorithm for optimal winner determination in combinatorial
  auctions,'' \emph{Artificial Intelligence}, vol. 135, no.~12, pp. 1--54, Feb.
  2002.

\bibitem{Wang2012}
F.~Wang, C.~Xu, L.~Song, Z.~Han, and B.~Zhang, ``Energy-efficient radio
  resource and power allocation for device-to-device communication underlaying
  cellular networks,'' in \emph{Proc. International Conference on Wireless
  Communications and Signal Processing}, Huangshan, China, Oct. 2012.

\bibitem{Saraydar2002}
C.~U. Saraydar, N.~B. Mandayam, and D.~J. Goodman, ``Efficient power control
  via pricing in wireless data networks,'' \emph{{IEEE} Trans. Commun.},
  vol.~50, no.~2, pp. 291--303, Feb. 2002.

\bibitem{Meshkati2006}
F.~Meshkati, M.~Chiang, H.~V. Poor, and S.~C. Schwartz, ``A game-theoretic
  approach to energy-efficient power control in multicarrier {CDMA} systems,''
  \emph{{IEEE} J. Sel. Areas Commun.}, vol.~24, no.~6, pp. 1115--1129, Jun.
  2006.

\bibitem{Wang2013a}
F.~Wang, C.~Xu, L.~Song, Q.~Zhao, X.~Wang, and Z.~Han, ``Energy-aware resource
  allocation for device-to-device underlay communication,'' in \emph{Proc. IEEE
  International Conference on Communications}, Budapest, Hungary, Jun. 2013,
  pp. 4669--4673.

\bibitem{Cui2004}
S.~Cui, A.~J. Goldsmith, and A.~Bahai, ``Energy-efficiency of {MIMO} and
  cooperative {MIMO} techniques in sensor networks,'' \emph{{IEEE} J. Sel.
  Areas Commun.}, vol.~22, no.~6, pp. 1089--1098, Aug. 2004.

\bibitem{Osborne1994}
M.~J. Osborne and A.~Rubinstein, \emph{A Course in Game Theory}.\hskip 1em plus
  0.5em minus 0.4em\relax MIT Press, Cambridge, MA, 1994.

\bibitem{Burden2011}
R.~Burden and J.~Faires, \emph{Numerical Analysis}, 9th~ed.\hskip 1em plus
  0.5em minus 0.4em\relax Brooks/Cole, 2011.

\bibitem{Conti2007}
A.~Conti, M.~Z. Win, and M.~Chiani, ``Slow adaptive {M-QAM} with diversity in
  fast fading and shadowing,'' \emph{{IEEE} Trans. Commun.}, vol.~55, no.~5,
  pp. 895--905, May 2007.

\bibitem{Li2010}
W.~W.-L. Li, Y.~Zhang, A.~M.-C. So, and M.~Z. Win, ``Slow adaptive {OFDMA}
  systems through chance constrained programming,'' \emph{{IEEE} Trans. Signal
  Process.}, vol.~58, no.~7, pp. 3858--3869, Jul. 2010.

\bibitem{Wang2013}
F.~Wang, L.~Song, Z.~Han, Q.~Zhao, and X.~Wang, ``Joint scheduling and resource
  allocation for device-to-device underlay communication,'' in \emph{Proc. IEEE
  Wireless Communications and Networking Conference}, Shanghai, China, Apr.
  2013, pp. 140--145.

\end{thebibliography}
\end{document}